\documentclass[prd,aps,preprint,tightenlines,nofootinbib,superscriptaddress]{revtex4}
\usepackage{mathrsfs}
\usepackage{amsfonts}
\usepackage{amsmath}
\usepackage{slashed}
\usepackage{array}
\usepackage{verbatim}
\usepackage{epsfig}
\usepackage{graphicx}
\usepackage{color}

\newcommand{\ep}{\epsilon}

\newcommand{\beq}{\begin{eqnarray}}
\newcommand{\eeq}{\end{eqnarray}}

\begin{document}

\title{More On Large-Momentum Effective Theory \\ Approach to Parton Physics}

\author{Xiangdong Ji}
\affiliation{Tsung-Dao Lee Institute, and College of Physics and Astronomy, Shanghai Jiao Tong University, Shanghai, 200240, P. R. China}
\affiliation{Department of Physics, University of Maryland, College Park, Maryland 20742, USA}
\author{Jian-Hui Zhang}
\affiliation{Institut f\"{u}r Theoretische Physik, Universit\"{a}t Regensburg, D-93040 Regensburg, Germany}
\author{Yong Zhao}
\affiliation{Center for Theoretical Physics, \\Massachusetts Institute of Technology, Cambridge, MA 02139, USA}

\preprint{MIT-CTP/4914}

\begin{abstract}

Large-Momentum Effective Theory (or LaMET) advocated by the present authors provides a direct
approach to simulate parton physics in Eulidean lattice QCD theory. Recently, there has been much interest in this theory
in the literature, with some questioning its validity and
effectiveness. Here we provide some discussions aiming at a further exposition of this approach.
In particular, we explain why it does not have the usual power divergence
problem in lattice QCD calculations for the moments of parton distributions. The only power divergence
in the LaMET approach comes from the self-energy of the Wilson lines which can be properly factorized. 
We show that although the Ioffe-time distribution provides an alternative way to extract
the parton distribution from the same lattice observables, it also requires the same large momentum 
(or short distance) limit as in LaMET to obtain a precision calculation. With a proper quantification of errors, 
both extraction methods shall be compared with the same lattice data.

\end{abstract}

\maketitle

\section{Introduction}
\label{intro}

The parton model invented by Feynman has been one of the most useful languages in describing the physics
of strong interactions at high energy~\cite{Feynman:1969ej}. Partons are idealized objects that exist when
hadrons travel at the speed of light. Thus in modern language, they are effective degrees of
freedoms in an effective theory, for instance, the soft-collinear
effective theory~\cite{Bauer:2000yr}.
The parton properties of a hadron, such as parton distributions, parton
wave functions or amplitudes, parton correlations, etc., are low-energy properties which
require solutions of quantum chromodynamics (QCD) in the strong coupling region.

When formulated in the rest frame of the hadron, parton physics represents
the space-time correlations of QCD quarks or gluons along the light-cone direction ($\xi^-=(t-z)/\sqrt{2}$),
which involves an explicit time dependence~\cite{Sterman:1994ce}.
(We use $\xi^\mu = (t, x, y, z)$ $(\mu=0,1,2,3)$ to denote space-time coordinates, although $x$, $y$, and $z$ are commonly used to
denote momentum fractions as well.)
In physics, the time-dependent
correlations are in a class by themselves because the time evolution calls for the presence of the
interaction-dependent Hamiltonian $H$, and the observables are called ``dynamical".

Dynamical correlations are hard to calculate for at least two different reasons: 1)
if one inserts a set of intermediate Hamiltonian eigenstates to simplify the evolution,
one needs to know the wave functions of
all the intermediate excited states. Thus knowing just the initial (often ground) state wave function is not
sufficient to calculate the time-dependent correlations. Methods designed
to approximate the ground state might not be applicable to the excited states, and thus dynamical
correlations are usually calculated less reliably than time-independent or static
observables for which the initial state wave function is sufficient. 2) Since the time-dependence
involves the unitary evolution operator, $e^{-iHt}$, it is a problem for numerical Monte Carlo simulations as
it contains a sign-changing phase. In fact, the majority of the Monte Carlo approaches designed
to calculate static properties cannot be readily applied to dynamical correlations. There have been attempts
to calculate instead an imaginary or Euclidean evolution with $e^{-H\tau}$, and to analytically continue
the result to the real time~\cite{Nakahara:1999vy,Asakawa:2000tr}. However,
these approaches are hardly reliable because
the analytical functions are unknown. An alternative trick is to Taylor-expand the
time-evolution operator, obtaining the time dependence through the sum of an infinite series.
Such an approach again is very limited because of the poor precision of calculations
for higher moments~\cite{Hagler:2009ni}. For the above reasons, the time-dependent correlations
are generally referred to as ``Minkowskian" in the sense that time and space correlations
are intrinsically different. On the other hand, the time-independent correlations may
be called ``Euclidean" , because they can be readily calculated with Monte Carlo simulations,
except that there may be additional sign problems.

For Minkowskian observables, the traditional approach is to use analytical approximations.
For instance, the so-called Schwinger-Dyson approach solves the time dependence
in terms of dynamical equations~\cite{Roberts:1994dr}. For parton physics, in the so-called light-front quantization approach,
the coordinates are redefined such that the new time is a combination of space and the old time, and
the light-cone correlations become time-independent~\cite{Burkardt:1995ct}. However, the resulting theory cannot be
solved readily using Monte Carlo methods because the metric is not positive definite.

Fortunately, there are certain ``dynamical" problems that are not intrinsically ``Minkowskian" in
the sense that with some clever formulation, they can be solved using Monte Carlo methods. For instance, scattering phase-shifts usually involve
complex S-matrix elements which cannot be calculated directly using a Monte Carlo approach.
However, one can put the theory in a box, calculating the bound state levels as a function of the boundary condition and extracting the phase shifts from bound state energy levels~\cite{Luscher:1986pf,Luscher:1990ck,Luscher:1990ux,Briceno:2017max}. A second example is certain Minkowskian correlators can be directly analytically continued to
Euclidean region without encountering any ``poles", and therefore,
can be calculated directly in Euclidean space~\cite{Ji:2001wha}.

Recently it was suggested by the authors that parton physics in non-perturbative QCD belongs to a class
of problems that are not intrinsically Minkowskian~\cite{Ji:2013fga,Ji:2013dva,Hatta:2013gta,Xiong:2013bka,
	Ji:2014gla,Ji:2014lra,Ji:2015jwa,Ji:2015qla,Xiong:2015nua}. This observation is based on two important
points: 1) light-cone correlators can be approximated by near-light-cone correlators through
QCD perturbation theory, thanks to asymptotic freedom. 2) the Lorentz boost is dynamical,
and it can be used to absorb the dynamical evolution between the fields through
boosting a hadron state. Therefore, all remainder dynamics is included in the boosted hadron wave function,
not in the probe operators. In this way, all parton physics can be accessed from equal-time correlations in a large-momentum hadron 
state which are calculable in Euclidean Monte Carlo simulations.

The approach is an effective field theory in the following sense.
For any time-dependent light-cone operator $O(\xi^-_1,\xi^-_2, ...)$,
one can construct an infinite number of Euclidean quasi-operators
${\cal O}_i(z_1,z_2,..)$, where $i$ is a label for all such operators, and $z_i$ are the spatial coordinates
of the fields along the direction of hadron momentum, which is taken as the third coordinate direction.
The matrix element of such an operator in a hadron state with a large momentum $p_z$, calculated for example in lattice QCD,  has a large momentum expansion~\cite{Ji:2014gla}
\begin{equation} \label{eq:fact}
\langle p|{\cal O}_i(z_1,z_2,..)|p\rangle
= C_{i0}(\alpha_s,p_z)\otimes\langle p|O(\xi^-_1,\xi^-_2, ...)|p\rangle + {\cal O}(1/p^2_z)\ ,
\end{equation}
where we have omitted all the renomalization scales, and $C_{i0}$ is a hard coefficient. The higher-order
corrections can be calculated systematically.
Since in this effective theory the large momentum is used to approach the light-cone, it is called the large momentum effective field theory
or LaMET for short.

It is worthwhile to point out that the choice of the Euclidean quasi-operator above is not unique. Any quasi-operator that yields the same light-front physics in the infinite momentum limit can be used to probe the desired light-cone operator. All such quasi-operators form a universality class. In practice, the existence of the universality class allows one to make an optimized choice of the quasi-operators so that the ${\cal O}(1/p_z^2)$ correction can be made as small as possible. One such example is the calculation of the total gluon polarization
in the nucleon~\cite{Hatta:2013gta}.

Recently, there has been much interest about this theory
in the literature, with some questioning its validity or
effectiveness. Here we provide some discussions aiming at a further exposition of this approach.
In Ref.~\cite{Carlson:2017gpk}, the factorization within the LaMET approach has been questioned. Following Ref.~\cite{Briceno:2017cpo},
we explicitly show that the problem arises from the procedure of analytical continuation. In Ref.~\cite{Rossi:2017muf},
the issue of power divergences has been raised based on an analysis of moments. We show that this problem does not exist in the non-local approach
where the only power divergence arises from the self-energy of the Wilson lines.
Finally, in Refs.~\cite{Radyushkin:2017cyf,Orginos:2017kos}, a new approach has been proposed to extract
parton distributions.  We show that the Ioffe-time
distribution is an alternative way to extract
parton distribution from the same lattice observable, and hence is entirely equivalent to the LaMET approach.
Besides, the apparent scaling observed in that approach might or might not help with the convergence problem.

\section{Perturbative Matching in Minkowski Space and Non-perturbative Renormalization in Euclidean Space}

In a recent paper~\cite{Carlson:2017gpk}, the validity about the factorization of
the quasi-distribution into parton distribution as in Eq. (1) has been questioned.
In the original one-loop factorization
proof in Ref.~\cite{Xiong:2013bka}, the quasi and physical parton distributions were
calculated using the Lehmann-Symanzik-Zimmermann reduction formula in Minkowskian field theory~\cite{Sterman:1994ce} where all the internal loop integrals are Minkowskian.
In this form, all the infrared (IR) properties are very clear. On the other hand, because the quasi-distribution
has no time-dependence, it can be calculated from a Euclidean space state that evolves as
\begin{equation}
|p\rangle = e^{-\beta H}|p)
\end{equation}
in Monte Carlo simulations. Here $|p\rm{)}$ is a state with momentum $p$ and correct quantum number but not an eigenstate, whereas $H$ is the QCD Hamiltonian, $\beta$ is an evolution parameter (``Euclidean time"). In lattice QCD, nucleon matrix elements are obtained from the source-operator-sink correlation where the source, operator, and sink are separated by imaginary time. For large separation, this three-point correlation function is dominated by the contribution from the lowest physical state~\cite{Rothe:1992nt}.
Thus, the matrix elements are effectively calculated in an Euclidean-like field theory. However,
the equivalence of the Euclidean calculations with the original Minkowskian field theory is guaranteed
through the derivation of the formalism. This equivalence has been the basis for all the
Euclidean lattice calculations in the past.

It was shown, however, that the Feynman integrals in lattice perturbation theory
with external Minkowskian momenta do not have collinear divergences~\cite{Carlson:2017gpk}. Therefore, it appears that the quasi-distribtutions from lattice QCD do not have proper IR properties.
In a subsequent paper~\cite{Briceno:2017cpo}, it was explicitly shown that the lattice
calculation procedure leads to the Minkowskian theory matrix element with correct collinear divergences.
The question then becomes what the correct procedure is to recover the physical
result (or collinear divergences) with Minkowskian external momenta from the
Euclidean perturbation theory. The answer is that one cannot directly insert
the Minskowskian external momentum into the Euclidean integrals, as this procedure will
change the pole structures. The correct procedure is to finish the Euclidean
integral with Euclidean external momenta and then analytically continue the result to Minkowski spacetime.
To see that the nucleon matrix elements of the quasi-distribution calculated from lattice QCD does capture the correct IR physics in perturbation theory, one has to perform an analytical continuation of the
matrix elements after the Euclidean loop integrals are done, not before.

In the following, we provide an example to illustrate the above point: We calculate a one-loop integral appearing in the Euclidean quasi-distribution and show that the collinear divergence becomes manifest after an analytic continuation from $p_E^2$ to $p_M^2$ with $p_M^\mu$ and $p_E^\mu$ being the Minkowskian and Euclidean momenta. Actually the result is identical to its Minkowskian counterpart.

We first start with the Minkowskian integral, where $p_M^0 = \sqrt{p_z^2+m^2}$,
\beq \label{eq:mloop}
I_M = \int {d^4k_M\over (2\pi)^4}{(2\pi)p^z\delta(k^z_M-xp^z)\over (k_M^2-m^2+i\epsilon) \left[(p_M-k_M)^2+i\epsilon\right]}\ ,
\eeq
We focus on the physical region $0<x<1$ where the collinear divergence could exist.
For $I_M$, we first integrate over $k_M^0$ with the residue theorem, and then over the transverse components, to get the result
\begin{align} 	\label{eq: im}
I_M =& {i\over 8\pi}{1\over \sqrt{1+\rho_M}}\left[\ln{1+\sqrt{1+\rho_M}\over \rho_M}\right.\nonumber\\
&\left. +\ln{\sqrt{(1+\rho_M)(x^2+\rho_M)}+x+\rho_M\over 1-x}\right]\ ,
\end{align}
where
\beq
\rho_M = {m^2\over (p^z)^2}<1\ .
\eeq

For the Euclidean space integral, we first keep $p^\tau_E$ real,
\beq \label{eq:eloop}
I_E = \int {d^4k_E\over (2\pi)^4}{(2\pi)p^z\delta(k^z_E-xp^z)\over (k_E^2+m^2) (p_E-k_E)^2}  \ .
\eeq
Integrating over the poles, we find
\begin{align}	\label{eq: ie}
I_E =& {1\over 8\pi\sqrt{\rho_E-1-i\ep}} \left[\arctan{\rho_E+\rho_M-2(1-x)\over2 \sqrt{\rho_E-1-i\ep}(1-x)} \right.\nonumber\\
&\left.+ \arctan {\rho_E-\rho_M-2x\over2 \sqrt{(\rho_E-1-i\ep)(x^2+\rho_M)}}\right] \ ,
\end{align}
where
\beq \label{eq: rho}
\rho_E={(p^\tau)^2+(p^z)^2\over (p^z)^2}>1\ .
\eeq
We keep $-i\ep$ in the imaginary part of $\rho_E$ so that we can analytically continue it to $\rho_M$ through the lower half complex plane by making the replacement $p^\tau \rightarrow -ip^0$,
\beq \label{eq: cont}
\sqrt{\rho_E-1-i\ep} \to -i\sqrt{1+\rho_M} \ .
\eeq
We find that
\begin{align} \label{eq: iecont}
I'_E =&-{1\over8\pi}{1\over \sqrt{1+\rho_M}} \left[ {1\over2}\text{arctanh}\left(- {1\over\sqrt{1+\rho_M}}\right)\right.\nonumber\\
&\left. + {1\over2}\text{arctanh}\left(-{x+\rho\over\sqrt{(1+\rho_M)(x^2+\rho_M)}}\right)\right]\nonumber\\
=& -i I_M \ ,
\end{align}
which is exactly the Minkowskian matrix element with a factor of $i$ that is to be canceled in $dk^0_M=idk^4_E$.
If we take the massless limit $m\to0$, then both $I_M$ and $I'_E$ give
\beq \label{eq: oslimit}
I_M = {i\over 8\pi} \left(\ln {(p^z)^2\over m^2} + \ln {4x\over1-x} \right) \ ,
\eeq
with collinear divergence correctly characterized by $\ln m^2$.

Despite the analytical continuation to recover the correct collinear divergence, we find that the Euclidean matrix elements with Euclidean external momenta are still useful for renormalization. The quasi-distribution calculated in lattice QCD requires a non-perturbative renormalization, which can be done, e.g., in the RI/MOM scheme used in Refs.~\cite{Alexandrou:2017huk,Chen:2017mzz,Stewart:2017tvs}. To compute the renormalization factors in such a scheme, one needs to compute the vertex functions with external quarks and gluons that have deep-Euclidean momenta. After renormalization, the quasi-distribution is expected to have a well-defined continuum limit. It can then be matched perturbatively to the normal distributions in the $\overline{\rm MS}$ scheme, where the matching factor can be computed in Minkowski space.

\section{Power Divergences and Renormalization of Quasi-Distributions in Lattice QCD}

In a paper by Rossi and Resta~\cite{Rossi:2017muf}, a question has been raised about the practical implementation
of the factorization theorem, Eq.~(\ref{eq:fact}), in lattice QCD. In particular, they consider
the quasi-distribution in Eq. (19) of their paper, and try to
compute its second moment in the rest frame of the hadron. They point out that
because of the mixing of the trace term with the lower dimensional operator, the
$1/p_z^2$ suppressed terms have a power-divergent contribution as shown in Eq. (35) of their paper.
Therefore, they conclude that ``unless one performs a non-perturbative subtraction of
the power divergent terms, the $p_z\rightarrow \infty$ limit does not exist. "
However, a careful analysis shows that this problem does not exist in the present formalism.

First of all, the quasi-distribution used to extract the physical
parton distribution, is a function of $y=k_z/p_z$. Formally, one can take the second
moment and show that it results in a local operator with double derivatives.
However, in practice, the moments of the quasi-distribution do not exist. As can be seen from the one-loop calculation~\cite{Xiong:2013bka},
the result behaves like $1/|y|$ at large $y$, therefore all higher
moments other than the zeroth one do not converge.
It is exactly this large $y$ divergence that calls for an extra renormalization
in the local operators. However, without taking the moments, the divergence
does not appear.

In fact, it is exactly this non-local quasi-distribution formulation of parton physics that
avoids the power divergence problem in moment calculations using the old-fashioned
approach. The non-local formulation has much simpler ultraviolet (UV) physics
and hence makes it much easier to control the divergences.

For example, the renormalization of the quasi-distributions is straightforward in ``heavy-quark" formalism. It can be shown
that for quark non-singlets, except for the Wilson-line self-energy or mass correction,
there are no other power divergences~\cite{Ishikawa:2016znu,Chen:2016fxx,Ji:2017oey,Ishikawa:2017faj,Green:2017xeu}. All other divergences are logarithmic
in the lattice spacing. Therefore, the quasi-distributions can be
renormalized by factoring out the power-divergent self-energy factor and logarithmic 
renormalization factor, and have a smooth limit when the lattice spacing vanishes.  
Note that without factoring out the power-divergent self-energy factor, 
a one-loop mass correction of form $\alpha_s C_1|z|/a$ with lattice spacing $a$
and Wilson-line length $|z|$ can generate a finite collinearly divergent term
in the quasi-distribution at two-loop, which appears to break the IR factorization
of the quasi-distribution~\cite{Li:2016amo}. However, such terms do not 
contribute when the linear divergence is factored out of the quasi-distributions before taking 
the lattice spacing to zero. 

After renormalizing the quasi-distributions in an appropriate scheme, one can then use the factorization formula in Eq. (\ref{eq:fact}) to match it to the $\overline{\text{MS}}$ parton distribution. Then the higher-twist $1/p_z^2$ terms vanish smoothly as $p_z\rightarrow \infty$.
This is exactly what RI/MOM scheme does in recent papers~\cite{Alexandrou:2017huk,Chen:2017mzz,Stewart:2017tvs}.

\section{Quasi-Distributions vs. Pseudo Distributions, Similarities and Differences}

In a recent paper~\cite{Radyushkin:2017cyf}, an alternative interpretation has been
suggested for the Euclidean matrix elements proposed in LaMET, which provides
another method to extract the physical parton distribution from the same lattice calculations.
The starting point is the so-called Ioffe-time distribution ${\cal M}(\nu,\xi^2)$,
which is defined as the matrix elements of bilinear quark operators, separated with arbitrary
spacetime distances $\xi^\mu$, and in a hadron state with four-momentum $p^\mu$. As such, it
is an invariant quantity of $\nu= -p\cdot \xi$ (Ioffe-time), and space-time invariant $-\xi^2$.
In practice, such a Lorentz-invariant function can be obtained
on the Euclidean lattice
from the two-quark spatial correlation function,
\begin{equation}
h(zp_z,z^2,\Lambda) =\langle p|\overline{\psi}(z)\Gamma L(z,0)\psi(0)|p\rangle \ ,
\end{equation}
where $\Gamma$ is Dirac matrix, $L(z,0)$ the straight gauge link, and $\Lambda$
a renormalization scale or lattice cutoff scale. Thus, the matrix element
$h(zp_z,z^2,\Lambda)$ in the quasi distribution calculations in LaMET
can be viewed as a covariant Ioffe-time distribution $h(\nu,\xi^2,\Lambda)$ with $\nu = zp_z$, in space-like separation, $-\xi^2 = z^2$.

Radyushkin suggests to interpret the Ioffe-time distribution in a new frame, in which
$\xi^\mu = (\xi^+=0, \xi^-/\gamma, \xi_\perp)$, where we have used the light-cone
variable $\xi^\mu = (\xi^+,\xi^-,\xi_\perp)$,  and $ p^\mu = (p^+\gamma, 0, p^-=M^2/(p^+\gamma)$,
$M$ is the hadron mass, and $\gamma$ is a boost parameter.
With $\gamma\rightarrow\infty$, one recovers the infinite momentum frame.
In this new frame, the Ioffe time is $\nu = -p^+\xi^-$, and $-\xi^2= \xi_\perp^2$ is again space-like,
and they are $\gamma$ independent. And $h$ becomes a near light-cone correlation function,
\begin{equation}\label{eq:lccorrel}
h(p^+\xi^-, \xi_\perp, \Lambda) =\langle p|\overline{\psi}(0,\xi^-,\xi_\perp)\Gamma L(0,\xi^-,\xi_\perp;0)\psi(0)|p\rangle,
\end{equation}
where now the gauge link has both components along the light-cone and transverse
directions. Lorentz symmetry guarantees the quality of Eq. (11) and (12), with $\xi_\perp = z$,
and $\xi^- = -zp_z/p^+$. The Fourier transformation with respect to the Ioffe-time $\nu$ has the direct
interpretation as the light-cone momentum fraction $x$, therefore is bounded in $[-1,1]$
as always the case in the infinite momentum frame.

One can then introduce a ``pseudo"-distribution from the space-like correlation,
\begin{equation}
{\cal P}(y, \xi_\perp, \Lambda) = \frac{1}{2\pi} \int^\infty_{-\infty}d\nu e^{-iy\nu}　h(zp_z,z^2, \Lambda)\ ,
\end{equation}
where $\xi_\perp^2 = z^2$. In this approach, the hadron momentum on lattice
is re-interpreted as the Ioffe time at a fixed $\xi_\perp$.
The physics of the pseudo-distribution is that it is a Fourier transformation
of a type of transverse-momentum dependent parton distributions. Since there
is a gauge link between the quark fields going along the transverse direction
in Eq.~(\ref{eq:lccorrel}), in $A^+=0$ guage, there is an infinite number of transversely polarized
gluons involved in the distributions. Therefore the parton picture of the pseudo-distribution
is not simple. In particular, the momentum fraction $y$ contains not only the contribution 
from a single quark, but also that from transversely-polarized physical gluons. $y$ approaches
the quark light-cone fraction only in the limit of $\xi_\perp=0$. 

In this alternative interpretation of the Euclidean matrix element,
the physical quark distribution is recovered by putting the quarks entirely on the light-cone,
i.e., when the transverse coordinate of the quark $\xi_\perp = z$ goes to zero.
In this limit, to get finite Ioffe-time $\nu = zp_z$, one again has to let
$p_z \rightarrow\infty$, the same as used in LaMET. Thus, the requirement to
get a precision parton distribution is exactly the same in the two approaches.

In practice, one has to work with small but nonvanishing $\xi_\perp$ to avoid light-cone divergences. 
One can then establish a similar factorization theorem as in LaMET (``a small-distance effective theory"),
\begin{equation}
{\cal P}(y, \xi_\perp,\tilde{\mu}) = \int \frac{dx}{x}
C(y/x, \xi_\perp\tilde{\mu},\tilde{\mu}/\mu) q(x, \mu) + {\cal O}(\xi^2_\perp),
\label{fac2}
\end{equation}
where $q$ is a physical parton distribution, $C$ is a matching coefficient that depends
on the specific lattice regularization, $\tilde{\mu}$ and $\mu$ are the renormalization scales of the Ioffe-time distribution and physical PDF.
To have continuum limit, $\xi_\perp$ shall be much
larger than lattice spacing $a$, but much smaller than $1/\Lambda_{\text{QCD}}$. The corrections are in terms of small $\xi_\perp$ expansion.
It would be interesting to extract the parton distributions through the above
factorization formula at a fixed $\xi_\perp$.

To demonstrate the above factorization theorem, let us take the isovector unpolarized quark distribution as an example. In the Feynman gauge, we calculate the quark matrix elements of the quasi and physical PDFs at one-loop order in dimensional regularization with $D=4-2\epsilon$. The external quark state is chosen to be onshell and massless, so the UV and collinear divergences are regularized by $1/\epsilon_{UV}\ (\epsilon_{UV}>0)$ and $1/\epsilon_{IR}\ (\epsilon_{IR}<0)$ respectively. We compute the same Feynman diagrams in Ref.~\cite{Xiong:2013bka} in the coordinate space to obtain $h^{(1)}(zp^z,z^2,\mu,\epsilon)$ (for $\Gamma=\gamma^0$) and $q^{(1)}(p^+z^-,\mu,\epsilon)$, where $\mu$ can be regarded as the renormalization scale. The Feynman rules are not conventional as there is a Fourier transform of one loop momentum component in each of the irreducible diagrams, for instance,
\begin{align}
\mu^{2\epsilon}\int {d^d k\over (2\pi)^d} {1\over k^2(p-k)^2} e^{-i k^z z} &=  i\mu^{2\epsilon}\int_0^1 du \int {d^d k\over (2\pi)^d} {e^{-i k^z z}\over k_E^4} e^{ -i u p^z z}\ .
\end{align}
Using Schwinger parametrization, we can evaluate the above integral analytically,
\begin{align}
&i\mu^{2\epsilon}\int_0^1 du\int {d^d k\over (2\pi)^d} {e^{-i k^z z}\over k_E^4} e^{-i u p^z z}\nonumber\\
&= i\mu^{2\epsilon}\int_0^1 du\int_0^\infty d\alpha\ \alpha\int {d^d k\over (2\pi)^d} e^{-\alpha k_E^4-i k^z z} e^{-i u p^z z} \nonumber\\
&= {i\over 16\pi^2}(\pi z^2\mu^2)^{\epsilon_{IR}} \Gamma(-\epsilon_{IR})\int_0^1 du\  e^{-iu p^z z}\ .
\end{align}
The complete results for $h^{(1)}(zp^z,z^2,\mu,\epsilon)$ and $q^{(1)}(p^+z^-,\mu,\epsilon)$ are
\begin{align}
&h^{(1)}(zp^z,z^2,\mu,\epsilon)\nonumber\\
=&{\alpha_sC_F\over 2\pi}\int_0^1 du \left[\left(1-u+\left({2u\over 1-u}\right)_+\right)\left(-\ln(z^2 \mu^2) - {1\over \epsilon'_{IR}}\right)\right.\nonumber\\
&\ \ \ \ \ \ \ \ + (1-u)\Big]e^{-iu p^z z}\nonumber\\
& + {\alpha_sC_F\over 2\pi}(izp^z)\int_0^1 du \int_0^1dt\ (2-u) (-\ln t^2) e^{-i(1-tu) p^z z} \nonumber\\
&  +\left[(1+\delta Z_\psi)+ {\alpha_sC_F\over 2\pi}\left(2\ln(z^2\mu^2) + {2\over \epsilon'_{UV}}+2\right)\right] e^{-i p^z z} \ ,
\end{align}

\begin{align}
&q^{(1)}(p^+z^-,\mu,\epsilon)\nonumber\\
=&{\alpha_sC_F\over 2\pi}\int_0^1 du \left(1-u+\left({2u\over 1-u}\right)_+\right)\left({1\over \epsilon'_{UV}}- {1\over \epsilon'_{IR}}\right)e^{iu p^+ z^-}\nonumber\\
& + (1+\delta Z_\psi)e^{i p^+ z^-} \ ,
\end{align}
where $Z_\psi = 1+ \delta Z_\psi$ is the quark wavefunction renormalization constant, and we have made the substitutions $1/\epsilon'_{UV,IR} = 1/\epsilon_{UV,IR}+\gamma_E+\ln\pi$. To guarantee vector current conservation, $Z_\psi$ is given in the on-shell scheme,
\beq
Z_\psi = 1 - {\alpha_sC_F\over 2\pi}{1\over2}\left({1\over \epsilon'_{UV}}- {1\over \epsilon'_{IR}}\right) + O(\alpha_s^2)\ . 
\eeq
Then we Fourier transform the Ioffe time $p^z z$ or $p^+z^-$ into $x$, and obtain
\begin{align} \label{eq:quasipdf}
&{\cal P}^{(1)}(x,\xi_\perp,\mu,\epsilon)\nonumber\\
=&{\alpha_sC_F\over 2\pi}\left[\left({1+x^2\over 1-x}\right)_+\left(-\ln(\xi_\perp^2 \mu^2) - {1\over \epsilon'_{IR}}-1\right)  - \left(4\ln(1-x)\over 1-x\right)_+  \right.\nonumber\\
&+2(1-x)\Big]\theta(x)\theta(1-x)\nonumber\\
&  +\left[ 1+ {\alpha_sC_F\over 2\pi}\left({3\over2}\ln(\xi_\perp^2\mu^2) + {3\over2}{1\over \epsilon'_{UV}}+{3\over2}\right)\right]\delta(1-x) \ ,
\end{align}
whereas the corresponding light-cone PDF is
\begin{align} \label{eq:lcpdf}
q^{(1)}(x,\mu,\epsilon) &= \delta(1-x) + {\alpha_s\over 2\pi}\left({1+x^2\over 1-x}\right)_+\left({1\over\epsilon'_{UV}} - {1\over \epsilon'_{IR}}\right)\ .
\end{align}
The support in the pseudo distribution ${\cal P}^{(1)}(x,\xi_\perp,\mu,\epsilon)$ is restricted to $0<x<1$, as expected.
If we renormalize both Eq.~(\ref{eq:quasipdf}) and Eq.~(\ref{eq:lcpdf}) in the $\overline{\text{MS}}$ scheme, then the matching coefficient in Eq.~(\ref{fac2}) is read off as
\begin{align}
&C\left(y,\xi_\perp\mu\right)\nonumber\\
=& \left[1+{\alpha_sC_F\over 2\pi}\left({3\over2}\ln(\xi_\perp^2\mu^2)+{3\over2}\right)\right] \delta(1-y)\nonumber\\
& + {\alpha_sC_F\over 2\pi} \left[-\left({1+y^2\over 1-y}\right)_+\left(\ln(\xi_\perp^2 \mu^2) +1 \right)- \left(4\ln(1-y)\over 1-y\right)_+ \right.\nonumber\\
&+2(1-y)\Big]\theta(y)\theta(1-y)\ .
\end{align}
\noindent 
Similar factorization formula can be extended to the case of lattice regularization and non-perturbative renormalization.

Instead of using the factorization formula in Eq.~(\ref{fac2}) to extract parton distribution
from lattice matrix element calculated at a fixed $\xi_\perp$, Ref. \cite{Radyushkin:2017cyf}
focused on the possible approximate factorized property of the Ioffe-time distribution,
i.e. ${\cal M}(\nu, z) \sim  g(\nu)h(z)$. If so, $g(\nu) \sim {\cal M}(\nu, z)/h(z)$ will have approximate
scaling at different $z$. One can then get the small $z$
Ioffe-time distribution from a large $z$ calculation, which does not need a very large
momentum to produce a large Ioffe time. Indeed, an exploratory calculation on lattice
shows that this special factorization or scaling works quite well~\cite{Orginos:2017kos}.

While such a scaling or factorization behavior is interesting, just like parton-hadron duality
in deep-inelastic scattering~\cite{Ji:1994br} allowing the extraction of parton distributions at small $Q^2$, it is of limited use in actual precision calculations
of parton distributions. The scaling in the exploratory calculations
was observed in a limited kinematic domain with limited precision.
The parton distributions involve Ioffe time at all sizes, particularly small
$x$ distribution involves very large Ioffe time, where the scaling becomes much harder to test
and the present result is certainly incorrect. To get the correct
large Ioffe-time behavior, one needs calculation at small $\xi_\perp$ and larger nucleon momenta, the same
as required in LaMET. Moreover, the new $\xi_\perp$ dependence factorization must be corrected
for precision calculations. Since there is no systematic approach to correct for the violation
of this factorization, such a phenomenological approach
will not be a replacement for rigorous approaches as in Eq. (\ref{fac2}), where
corrections can be quantified.

It of course will be interesting to compare the parton distributions from two seemingly
different factorization approaches (large momentum vs. small distance), when extracted from the same lattice matrix elements.
It is important to quantify the errors of extraction in both approaches consistently.
We emphasize that LaMET is a much more general framework to calculate parton physics
with simple physical pictures as provided by Feynman. It has a well-defined recipe on
how to systematically calculate all parton observables with quantifiable errors,
including multi-parton amplitudes and correlations.

\section{Conclusion}

We have presented some discussions aiming at a further exposition of the LaMET approach.
We clarified the validity of the factorization in this approach, and stressed the importance
of analytical continuation in recovering the correct IR behavior of Minkowski space integrals
from their Euclidean counterparts. We also pointed out that power divergences plaguing the traditional
moments approach do not pose a problem in the quasi-distribution approach, which employs
a non-local formulation.
Finally, we showed that the Ioffe-time distribution method to extract the parton
distribution from the same lattice observables used in LaMET requires exactly the same
physical conditions. It is interesting to compare the results of both approaches
after proper quantifications of errors.

\section*{Acknowledgement}
This work was partially supported by the U.S. Department of Energy Office of Science, Office of Nuclear Physics under Award Number DE-FG02-93ER-40762 and DE-SC0011090,
and the SFB/TRR-55 grant ``Hadron Physics from Lattice QCD". YZ was supported in part by the MIT MISTI program. The work of XJ and YZ was also supported in part by the
U.S. Department of Energy, Office of Science, Office of Nuclear Physics, within the framework of the TMD Topical Collaboration,
and a grant from Science and Technology Commission of Shanghai Municipality (Grants No. 16DZ2260200).


\begin{thebibliography}{00}
	
	\bibitem{Feynman:1969ej}
	R.~P.~Feynman,
	Phys.\ Rev.\ Lett.\  {\bf 23}, 1415 (1969).
	doi:10.1103/PhysRevLett.23.1415
	
	
	\bibitem{Bauer:2000yr}
	C.~W.~Bauer, S.~Fleming, D.~Pirjol and I.~W.~Stewart,
	Phys.\ Rev.\ D {\bf 63}, 114020 (2001)
	doi:10.1103/PhysRevD.63.114020
	[hep-ph/0011336].
	
	
	\bibitem{Sterman:1994ce}
	G.~F.~Sterman,
	``An Introduction to quantum field theory,"
	Cambridge, UK: University Press, 1993.
	
	
	\bibitem{Nakahara:1999vy}
	Y.~Nakahara, M.~Asakawa and T.~Hatsuda,
	Phys.\ Rev.\ D {\bf 60}, 091503 (1999)
	doi:10.1103/PhysRevD.60.091503
	[hep-lat/9905034].
	
	
	\bibitem{Asakawa:2000tr}
	M.~Asakawa, T.~Hatsuda and Y.~Nakahara,
	Prog.\ Part.\ Nucl.\ Phys.\  {\bf 46}, 459 (2001)
	doi:10.1016/S0146-6410(01)00150-8
	[hep-lat/0011040].
	
	
	\bibitem{Hagler:2009ni}
	P.~Hagler,
	Phys.\ Rept.\  {\bf 490}, 49 (2010)
	doi:10.1016/j.physrep.2009.12.008
	[arXiv:0912.5483 [hep-lat]].
	
	
	\bibitem{Roberts:1994dr}
	C.~D.~Roberts and A.~G.~Williams,
	Prog.\ Part.\ Nucl.\ Phys.\  {\bf 33}, 477 (1994)
	doi:10.1016/0146-6410(94)90049-3
	[hep-ph/9403224].
	
	
	\bibitem{Burkardt:1995ct}
	M.~Burkardt,
	Adv.\ Nucl.\ Phys.\  {\bf 23}, 1 (1996)
	doi:10.1007/0-306-47067-5\_1
	[hep-ph/9505259].
	
	
	\bibitem{Luscher:1986pf}
	M.~Luscher,
	Commun.\ Math.\ Phys.\  {\bf 105}, 153 (1986).
	doi:10.1007/BF01211097
	
	
	\bibitem{Luscher:1990ck}
	M.~Luscher and U.~Wolff,
	Nucl.\ Phys.\ B {\bf 339}, 222 (1990).
	doi:10.1016/0550-3213(90)90540-T
	
	
	\bibitem{Luscher:1990ux}
	M.~Luscher,
	Nucl.\ Phys.\ B {\bf 354}, 531 (1991).
	doi:10.1016/0550-3213(91)90366-6
	
	
	\bibitem{Briceno:2017max}
	R.~A.~Briceno, J.~J.~Dudek and R.~D.~Young,
	arXiv:1706.06223 [hep-lat].
	
	
	
	\bibitem{Ji:2001wha}
	X.~d.~Ji and C.~w.~Jung,
	Phys.\ Rev.\ Lett.\  {\bf 86}, 208 (2001)
	doi:10.1103/PhysRevLett.86.208
	[hep-lat/0101014].
	
	
	\bibitem{Ji:2013fga}
	X.~Ji, J.~H.~Zhang and Y.~Zhao,
	Phys.\ Rev.\ Lett.\  {\bf 111}, 112002 (2013)
	doi:10.1103/PhysRevLett.111.112002
	[arXiv:1304.6708 [hep-ph]].
	
	
	
	\bibitem{Ji:2013dva}
	X.~Ji,
	Phys.\ Rev.\ Lett.\  {\bf 110}, 262002 (2013)
	doi:10.1103/PhysRevLett.110.262002
	[arXiv:1305.1539 [hep-ph]].
	
	
	\bibitem{Hatta:2013gta}
	Y.~Hatta, X.~Ji and Y.~Zhao,
	Phys.\ Rev.\ D {\bf 89}, no. 8, 085030 (2014)
	doi:10.1103/PhysRevD.89.085030
	[arXiv:1310.4263 [hep-ph]].
	
	
	\bibitem{Xiong:2013bka}
	X.~Xiong, X.~Ji, J.~H.~Zhang and Y.~Zhao,
	Phys.\ Rev.\ D {\bf 90}, no. 1, 014051 (2014)
	doi:10.1103/PhysRevD.90.014051
	[arXiv:1310.7471 [hep-ph]].
	
	
	\bibitem{Ji:2014gla}
	X.~Ji,
	Sci.\ China Phys.\ Mech.\ Astron.\  {\bf 57}, 1407 (2014)
	doi:10.1007/s11433-014-5492-3
	[arXiv:1404.6680 [hep-ph]].
	
	
	\bibitem{Ji:2014lra}
	X.~Ji, J.~H.~Zhang and Y.~Zhao,
	Phys.\ Lett.\ B {\bf 743}, 180 (2015)
	doi:10.1016/j.physletb.2015.02.054
	[arXiv:1409.6329 [hep-ph]].
	
	
	\bibitem{Ji:2015jwa}
	X.~Ji and J.~H.~Zhang,
	Phys.\ Rev.\ D {\bf 92}, 034006 (2015)
	doi:10.1103/PhysRevD.92.034006
	[arXiv:1505.07699 [hep-ph]].
	
	
	\bibitem{Ji:2015qla}
	X.~Ji, A.~Schäfer, X.~Xiong and J.~H.~Zhang,
	Phys.\ Rev.\ D {\bf 92}, 014039 (2015)
	doi:10.1103/PhysRevD.92.014039
	[arXiv:1506.00248 [hep-ph]].
	
	
	\bibitem{Xiong:2015nua}
	X.~Xiong and J.~H.~Zhang,
	Phys.\ Rev.\ D {\bf 92}, no. 5, 054037 (2015)
	doi:10.1103/PhysRevD.92.054037
	[arXiv:1509.08016 [hep-ph]].
	
	
	\bibitem{Carlson:2017gpk}
	C.~E.~Carlson and M.~Freid,
	Phys.\ Rev.\ D {\bf 95}, no. 9, 094504 (2017)
	doi:10.1103/PhysRevD.95.094504
	[arXiv:1702.05775 [hep-ph]].
	
	
	\bibitem{Briceno:2017cpo}
	R.~A.~Briceño, M.~T.~Hansen and C.~J.~Monahan,
	arXiv:1703.06072 [hep-lat].
	
	
	\bibitem{Rossi:2017muf}
	G.~C.~Rossi and M.~Testa,
	arXiv:1706.04428 [hep-lat].
	
	
	\bibitem{Radyushkin:2017cyf}
	A.~V.~Radyushkin,
	arXiv:1705.01488 [hep-ph].
	
	
	\bibitem{Orginos:2017kos}
	K.~Orginos, A.~Radyushkin, J.~Karpie and S.~Zafeiropoulos,
	arXiv:1706.05373 [hep-ph].
	
	
	\bibitem{Rothe:1992nt}
	H.~J.~Rothe,
	``Lattice gauge theories: An Introduction,''
	World Sci.\ Lect.\ Notes Phys.\  {\bf 43}, 1 (1992)
	[World Sci.\ Lect.\ Notes Phys.\  {\bf 59}, 1 (1997)]
	[World Sci.\ Lect.\ Notes Phys.\  {\bf 74}, 1 (2005)]
	[World Sci.\ Lect.\ Notes Phys.\  {\bf 82}, 1 (2012)].
	
	
	
	
	
	
	
	\bibitem{Alexandrou:2017huk}
	C.~Alexandrou, K.~Cichy, M.~Constantinou, K.~Hadjiyiannakou, K.~Jansen, H.~Panagopoulos and F.~Steffens,
	arXiv:1706.00265 [hep-lat].
	
	
	\bibitem{Chen:2017mzz}
	J.~W.~Chen, T.~Ishikawa, L.~Jin, H.~W.~Lin, Y.~B.~Yang, J.~H.~Zhang and Y.~Zhao,
	arXiv:1706.01295 [hep-lat].
	
	
	
	\bibitem{Stewart:2017tvs} 
	I.~W.~Stewart and Y.~Zhao,
	arXiv:1709.04933 [hep-ph].
	
	
	\bibitem{Ishikawa:2016znu} 
	T.~Ishikawa, Y.~Q.~Ma, J.~W.~Qiu and S.~Yoshida,
	arXiv:1609.02018 [hep-lat].
	
	
	\bibitem{Chen:2016fxx} 
	J.~W.~Chen, X.~Ji and J.~H.~Zhang,
	Nucl.\ Phys.\ B {\bf 915}, 1 (2017)
	doi:10.1016/j.nuclphysb.2016.12.004
	[arXiv:1609.08102 [hep-ph]].
	
	
	\bibitem{Ji:2017oey} 
	X.~Ji, J.~H.~Zhang and Y.~Zhao,
	arXiv:1706.08962 [hep-ph].
	
	
	\bibitem{Ishikawa:2017faj} 
	T.~Ishikawa, Y.~Q.~Ma, J.~W.~Qiu and S.~Yoshida,
	arXiv:1707.03107 [hep-ph].
	
	\bibitem{Green:2017xeu} 
	J.~Green, K.~Jansen and F.~Steffens,
	arXiv:1707.07152 [hep-lat].
	
	
	\bibitem{Li:2016amo} 
	H.~n.~Li,
	Phys.\ Rev.\ D {\bf 94}, no. 7, 074036 (2016)
	doi:10.1103/PhysRevD.94.074036
	[arXiv:1602.07575 [hep-ph]].
	
	
	\bibitem{Ji:1994br}
	X.~D.~Ji and P.~Unrau,
	Phys.\ Rev.\ D {\bf 52}, 72 (1995)
	doi:10.1103/PhysRevD.52.72
	[hep-ph/9408317].
		
\end{thebibliography}
\end{document}